\documentclass[10pt]{JINST}

\usepackage{psfig}
\usepackage{graphicx}

\title{The Positioning System of the ANTARES Neutrino Telescope}

\author{S.~Adri\'an-Mart\'inez$^a$, M. Ageron$^c$, J.A. Aguilar$^b$, I. Al Samarai$^c$, A. Albert$^d$, M.~Andr\'e$^e$, M. Anghinolfi$^f$, G. Anton$^g$, S. Anvar$^h$, M. Ardid$^a$
% ...
\\
\llap{$^a$}Institut d'Investigaci\'o per a la Gesti\'o Integrada de les Zones Costaneres (IGIC) - Universitat Polit\`ecnica de Val\`encia, 
C/  Paranimf 1 , 46730 Gandia, Spain\\
\llap{$^b$}IFIC - Instituto de F\'isica Corpuscular, Edificios Investigaci\'on de Paterna, CSIC - Universitat de Val\`encia,  Apdo. de Correos 22085, 46071 Valencia, Spain\\
\llap{$^c$}CPPM, Aix-Marseille Universit\'e, CNRS/IN2P3, 
Marseille, France\\
\llap{$^d$}GRPHE - Institut universitaire de technologie de Colmar 34 rue du Grillenbreit BP 50568 - 68008 Colmar, France\\
\llap{$^e$}Technical University of Catalonia, Laboratory of Applied Bioacoustics, Rambla Exposici\'o,08800 Vilanova i la Geltr\'u,Barcelona, Spain\\
\llap{$^f$}INFN - Sezione di Genova, Via Dodecaneso 33, 16146 Genova, Italy\\
\llap{$^g$}Friedrich-Alexander-Universit\"at Erlangen-N\"urnberg, Erlangen Centre for Astroparticle Physics, Erwin-Rommel-Str. 1, 91058 Erlangen, Germany\\
\llap{$^h$}Direction des Sciences de la Mati\`ere - Institut de recherche sur les lois fondamentales de l'Univers - Service d'Electronique des D\'etecteurs et d'Informatique, CEA Saclay, 91191 Gif-sur-Yvette Cedex, France\\
\llap{$^i$}Nikhef, Science Park,  Amsterdam, The Netherlands\\
\llap{$^j$}APC - Laboratoire AstroParticule et Cosmologie, UMR 7164 (CNRS, Universit\'e Paris 7 Diderot, CEA, Observatoire de Paris) 10, rue Alice Domon et L\'eonie Duquet 75205 Paris Cedex 13,  France\\
\llap{$^k$}LAM - Laboratoire d'Astrophysique de Marseille, P\^ole de l'\'Etoile Site de Ch\^ateau-Gombert, rue Fr\'ed\'eric Joliot-Curie 38,  13388 Marseille Cedex 13, France\\
\llap{$^l$}INFN - Sezione di Bologna, Viale Berti-Pichat 6/2, 40127 Bologna, Italy\\
\llap{$^m$}Dipartimento di Fisica dell'Universit\`a, Viale Berti Pichat 6/2, 40127 Bologna, Italy\\
\llap{$^n$}INFN - Sezione di Pisa, Largo B. Pontecorvo 3, 56127 Pisa, Italy\\
\llap{$^o$}INFN -Sezione di Roma, P.le Aldo Moro 2, 00185 Roma, Italy\\
\llap{$^p$}Dipartimento di Fisica dell'Universit\`a La Sapienza, P.le Aldo Moro 2, 00185 Roma, Italy\\
\llap{$^q$}Clermont Universit\'e, Universit\'e Blaise Pascal, CNRS/IN2P3, Laboratoire de Physique Corpusculaire, BP 10448, 63000 Clermont-Ferrand, France\\
\llap{$^r$}G\'eoazur - Universit\'e de Nice Sophia-Antipolis, CNRS/INSU, IRD, Observatoire de la C\^ote d'Azur and Universit\'e Pierre et Marie Curie, BP 48, 06235 Villefranche-sur-mer, France\\
\llap{$^s$}INFN - Sezione di Bari, Via E. Orabona 4, 70126 Bari, Italy\\
\llap{$^t$}COM - Centre d'Oc\'eanologie de Marseille, CNRS/INSU et Universit\'e de la M\'editerran\'ee, 163 Avenue de Luminy, Case 901, 13288 Marseille Cedex 9, France\\
\llap{$^u$}INFN - Laboratori Nazionali del Sud (LNS), Via S. Sofia 62, 95123 Catania, Italy\\
\llap{$^v$}Univ Paris-Sud, 91405 Orsay Cedex, France\\
\llap{$^w$}Kernfysisch Versneller Instituut (KVI), University of Groningen, Zernikelaan 25, 9747 AA Groningen, The Netherlands\\
\llap{$^x$}Dipartimento di Fisica dell'Universit\`a, Largo B. Pontecorvo 3, 56127 Pisa, Italy\\
\llap{$^y$}Royal Netherlands Institute for Sea Research (NIOZ), Landsdiep 4,1797 SZ 't Horntje (Texel), The Netherlands\\
\llap{$^z$}Dr. Remeis-Sternwarte and ECAP, Universit\"at Erlangen-N\"urnberg,  Sternwartstr. 7, 96049 Bamberg, Germany\\
\llap{$^{aa}$}Universiteit Utrecht, Faculteit Betawetenschappen, Princetonplein 5, 3584 CC Utrecht, The Netherlands\\
\llap{$^{ab}$}Universiteit van Amsterdam, Instituut voor Hoge-Energie Fysica, Science Park 105, 1098 XG Amsterdam, The Netherlands\\
\llap{$^{ac}$}Moscow State University,Skobeltsyn Institute of Nuclear Physics, Leninskie gory, 119991 Moscow, Russia\\
\llap{$^{ad}$}INFN - Sezione di Catania, Viale Andrea Doria 6, 95125 Catania, Italy\\
\llap{$^{ae}$}Dipartimento di Fisica ed Astronomia dell'Universit\`a, Viale Andrea Doria 6, 95125 Catania, Italy\\
\llap{$^{af}$}Direction des Sciences de la Mati\`ere - Institut de recherche sur les lois fondamentales de l'Univers - Service de Physique des Particules, CEA Saclay, 91191 Gif-sur-Yvette Cedex, France\\
\llap{$^{ag}$}D\'epartement de physique nucl\'eaire et corpusculaire, Universit\'e de Gen\`eve, 1211, Switzerland\\
\llap{$^{ah}$}Institute for Space Sciences,\\ R-77125 Bucharest, M\u{a}gurele, Romania\\
\llap{$^{ai}$}IPHC-Institut Pluridisciplinaire Hubert Curien - Universit\'e de Strasbourg et CNRS/IN2P3  23 rue du Loess, BP 28,  67037 Strasbourg Cedex 2, France\\
\llap{$^{aj}$}ITEP - Institute for Theoretical and Experimental Physics, B. Cheremushkinskaya 25, 117218 Moscow, Russia\\
\llap{$^{ak}$}Dipartimento di Fisica dell'Universit\`a, Via Dodecaneso 33, 16146 Genova, Italy\\
E-mail: \email{juergen.hoessl@physik.uni-erlangen.de}
}

\abstract{The ANTARES neutrino telescope, located 40$\,$km off the coast of Toulon in the Mediterranean Sea at a mooring depth of about 2475$\,$m, consists of twelve detection  lines equipped typically with 25 storeys. Every storey carries three optical modules that detect Cherenkov light induced by charged secondary particles (typically muons) coming from neutrino interactions. As these lines are flexible structures fixed to the sea bed and held taut by a buoy, sea currents cause the lines to move and the storeys to rotate.  The knowledge of the position of the optical modules with a precision better than 10$\,$cm is essential for a good reconstruction of particle tracks. In this paper the ANTARES positioning system is described. It consists of an acoustic positioning system, for distance triangulation, and a compass-tiltmeter system, for the measurement of the orientation and inclination of the storeys. Necessary corrections are discussed and the results of the detector alignment procedure are described.}

\keywords{ANTARES neutrino telescope, detector alignment, acoustic positioning, calibration}

\begin{document}

\section{Introduction}

\begin{figure}[ht]
 \begin{center}
 \centerline{\psfig{figure=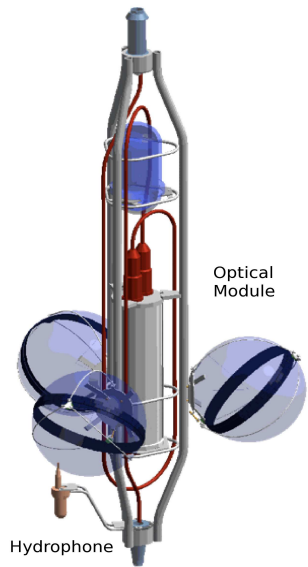,width=4.5cm,height=8.cm}}
 \end{center}
 \caption{Schematic diagram of an ANTARES storey carrying three optical modules, an electronics container and a positioning hydrophone fixed off-axis.}  \label{storey}
\end{figure}

\begin{figure}[ht]
 \begin{center}
  \psfig{figure=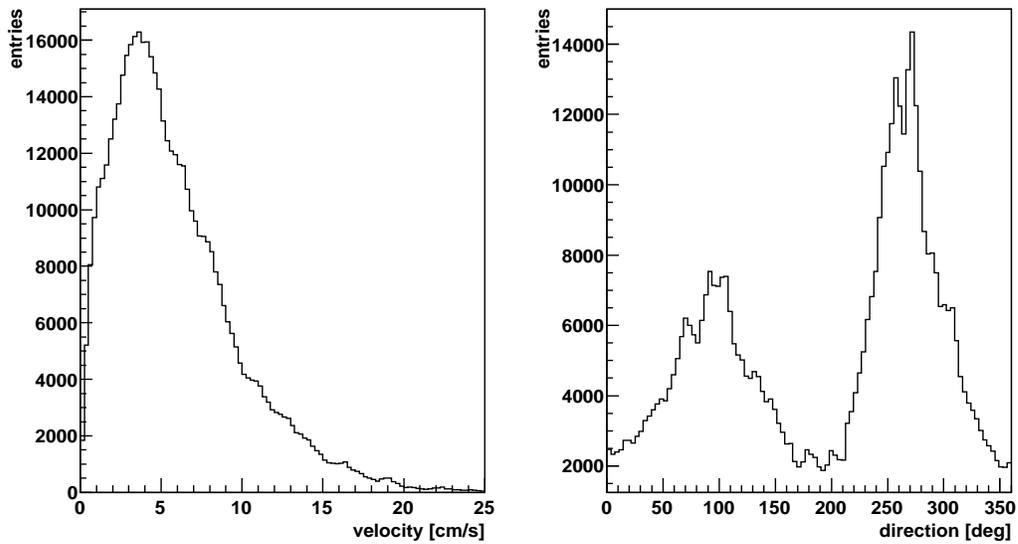,width=14.0cm,height=8.2cm}
 \end{center}
 \caption{Distributions of  sea current speed (left) and horizontal direction (right)  at the ANTARES site measured with an  Acoustic Doppler Current Profiler (ADCP) in 2009 \cite{detector}. The velocities are typically around 5$\,$cm/s. Two preferred directions due to the Ligurian current (East 90$^{\circ}$, West 270$^{\circ}$) are clearly visible.}
  \label{adcp}
\end{figure}

The ANTARES neutrino telescope \cite{detector,AMADEUS}, located in the Mediterranean Sea 40$\,$km off the coast of
Toulon, France at a mooring depth of about 2475$\,$m, is designed to search for high-energy cosmic neutrinos from astrophysical sources \cite{diffuse,point}. Neutrinos are measured by the detection of
Cherenkov light induced by charged secondary particles from neutrino
interactions occurring in the volume inside or surrounding the detector. The ANTARES detector consists of
 885 optical  modules arranged in a three dimensional array over
twelve detection lines. The inter-line spacing is 60-70$\,$m. Typically, the lines have a height of about
450$\,$m and carry 25 storeys. Each storey is a support structure for three optical
modules and an electronic container, see Figure \ref{storey}.
  The cables between the storeys
serve both as mechanical structure and as an electro-optical connection. 
The lines are fixed to the sea floor by an anchor called a 'bottom string socket' (BSS) and are held taut by
a buoy at the top of the line. The lines are not rigid
structures, so deep-sea currents (typically around 5$\,$cm/s, see Figure \ref{adcp}) cause the lines to be displaced from the vertical 
and the storeys to rotate around the line axis.

 The optical modules  \cite{OM}  consist of a pressure resistant glass sphere   housing a 10$''$ Hamamatsu photomultiplier (PMT) oriented 45$^{\circ}$ downwards in order to optimize the detection of light from upward going muons.
 The algorithm for track reconstruction in ANTARES is based on the measurement of the arrival times of the Cherenkov photons \cite{point,reco}. To achieve a direction resolution of a few tenths of a degree,  it is necessary to know the position of each optical module with a precision of about 10$\,$cm (corresponding to an uncertainty in the travel time of light in water of 0.5$\,$ns). This allows to fully exploit the timing resolution of about 1 ns for the recorded PMT signals \cite{time} without degradation by positioning uncertainties.
To achieve this accuracy during data taking, the ANTARES detector incorporates  a two-fold system.

The first is an acoustic positioning system consisting of  emitters on the sea floor around the detector and at the anchor of each line, as well as receiving hydrophones on five storeys of each line. This subsystem measures the position of the storeys by triangulation (see Section \ref{APS}).
The second subsystem consists of  a compass-tiltmeter system installed on each storey which measures the orientation and inclination of each storey (see Section \ref{CTS}).
 Information provided by  both systems is combined  into a global $\chi^2$-fit that reconstructs the shape of each line. The position and orientation of each optical module is then determined from the line shape as explained in Sections \ref{lssection} and \ref{LFIT}.

\section{The Acoustic Positioning System \label{APS}}

The acoustic positioning system of the ANTARES experiment is composed of two subsystems.
The low frequency long baseline positioning system (LFLBL)  is used to connect the  local frame of the detector to the geodetic reference frame. This allows the absolute position and orientation of the detector to be determined. The high frequency long baseline positioning system (HFLBL) serves to measure the relative positions of the different elements of the ANTARES detector within the local detector frame with high accuracy.

\subsection{The Low Frequency Long Baseline Positioning System \label{LFLBL}}

The LFLBL acoustic positioning system is a commercial set of devices provided by the IXSEA company\footnote{\it http://www.ixsea.com}. This system uses the 8-16$\,$kHz frequency range  to measure the travel time of controlled acoustic pulses between an emitter-receiver acoustic module (AM), linked to a time counting electronic system, and autonomous battery-powered transponders. The use of acoustic signals at these frequencies allows long propagation times to be measured. After convolution with the sound velocity profile of sea water, distances  up to about 8000$\,$m can be resolved. Therefore,  this system is well suited to determine the position of ANTARES structures  by triangulation from a ship on the sea surface.

In the ANTARES experiment, every line anchor (BSS) is equipped with an autonomous transponder, allowing its  positioning to be determined by triangulation with respect to a set of five reference acoustic transponders, placed on the sea bed around the detector area at distances of about 1100-1600$\,$m. In addition, an acoustic module is operated from a ship at a depth of 15-20$\,$m to avoid the interference of  acoustic reflections off the sea surface or warm water layers present at shallow depths during the summer season. The absolute geodetic positions of the reference transponders were individually determined with an accuracy better than 1$\,$m prior to the installation of the ANTARES detector. This was done by performing several hundred triangulations from a  ship positioned by Differential GPS (DGPS). The sound-velocity profile from the sea surface to the sea floor necessary for this calculation has been derived from the Chen-Millero formula \cite{schall} using sets of temperature and salinity profiles measured at the ANTARES site.

Combining the acoustic travel time measurements between the transponders on the BSS, the reference transponders and the AM on the ship, the absolute position of the  BSS is monitored in real time during the line deployment. The position is determined with an accuracy of a few meters and the final anchoring position of the BSS with an accuracy better than 1$\,$m after statistical averaging.

In order to determine the absolute orientation of the detector, accurate measurements of the BSS positions are vital.
Therefore, the relative positions of the line anchors are constrained using all distances between pairs of high frequency positioning transducers obtained from the HFLBL system (accuracy about 3$\,$cm, see Section \ref{high1}). Since the high frequency transponders are mounted about one metre off-axis with respect to the line -- whereas the LFLBL transponder is located close to the line axis -- the orientation of the BSS has to be considered. The latter is determined with an accuracy of about 5$^{\circ}$ using the compass of a submarine vehicle during an undersea line connection. In addition, the relative depth of the sea bed at the position of each BSS is measured by a pressure sensor on the submarine with a precision of about 10$\,$cm. Using the HFLBL triangulation data of periods where  the line remains almost vertical because of small speeds of the sea current  (smaller than 2$\,$cm/s) and  combining with the supplementary data described above, the determination of the BSS depth is substantially improved.  

The absolute orientation of the neutrino telescope with respect to the sky is obtained using both the absolute positions of the different line anchors and the BSS to BSS relative positions. 
An estimate of the accuracy of the absolute pointing of the ANTARES telescope was obtained by Monte Carlo techniques taking into account the accuracy of the individual BSS positions, BSS to BSS distances and the uncertainty of the sound velocity \cite{Gabed}. The resulting uncertainty for the absolute orientation was found to be less than 0.13$\,^{\circ}$ in the horizontal and  less than 0.06$\,^{\circ}$ in the vertical directions. This well matches the resolution of about 0.2$\,^{\circ}$ aimed at for high energy muon track reconstruction.

\subsection{The High Frequency Long Baseline Positioning System \label{high1}}

The HFLBL acoustic positioning system of ANTARES was  developed and constructed by the company GENISEA/ECA\footnote{Formerly GENISEA now ECA, {\it http://www.eca.fr}}. It is used to determine the position of the detector elements relative to the BSS positions with high precision. The positioning method is based on a measurement of travel times of acoustic sinusoidal pulses (40--60$\,$kHz) between  acoustic transceivers fixed at the line anchors and receiving hydrophones on the detector lines. These acoustic distance measurements are then combined to obtain the positions of the hydrophones by triangulation. For the frequencies used, the attenuation length in sea water of  700-1000$\,$m \cite{abs} is sufficiently long for acoustic measurements over the ANTARES dimensions.   

In addition to the battery powered transponder of the LFLBL, each ANTARES detector line is equipped with a transceiver  fixed on a rod at the BSS. It is operated as receiver and emitter (RxTx module). Five storeys of each line (storeys 1, 8, 14, 20 and 25, counted from below)  carry receiving hydrophones (Rx modules) to detect the signals of the emitters. 
The distances between the hydrophones are smaller in the upper part of the line where the displacement from the nominal position is larger.

The RxTx modules are composed of one transceiver and six electronic boards located in the String Control Module (SCM) of each BSS. Their purpose is to emit acoustic signals (to be detected by the Rx modules), triggered by an external synchronisation signal (Master Clock), to detect the signals of other RxTx modules, to stamp the detection time with respect to the Master Clock and to transmit  the timestamps and amplitudes to the shore station.
The Rx modules are composed of one hydrophone and three electronic cards in the Local Control Module (LCM) of  the respective storey. They fulfill the same tasks as the RxTx modules except the emission of acoustic signals.

\begin{figure}[ht]
 \begin{center}
 \centerline{\psfig{figure=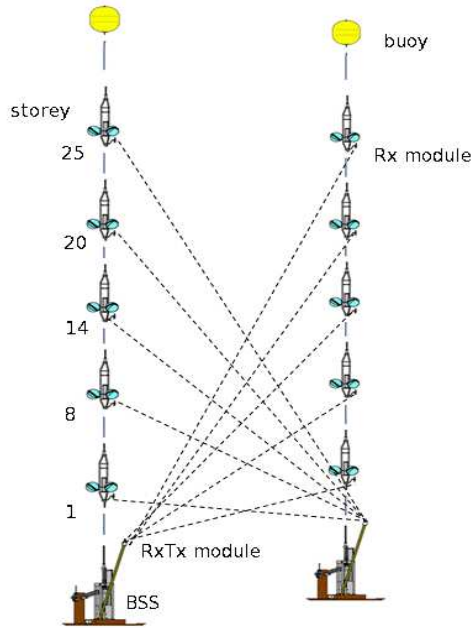,width=8.5cm,height=8.5cm}}
 \end{center}
 \caption{Schematic demonstrating the principle of the HFLBL positioning system for two lines (for simplicity only storeys with an Rx module are shown).}  \label{schematic}
\end{figure}

The positioning of the ANTARES detector is done by successively sending acoustic wavepackets (typical duration 2$\,$ms) from each RxTx module at  different frequencies  to identify unambiguously the emitting module. The acoustic emission is organized in a periodic succession of cycles (every two minutes, programmable from onshore). One cycle consists of the emissions of all RxTx modules at their respective frequencies between 44.522$\,$kHz and 60.235$\,$kHz in a predefined sequence.  Following the emission of a specific RxTx module  all other RxTx and Rx modules are put into reception mode, searching for the emitted frequency with a band-pass filter. 
Once the modules detect  acoustic signals at the given frequency above a predefined  threshold they provide the detection time together with the measured amplitude of the signal. The threshold as well as the amplification gain of every receiving module can be set individually from the shore. This allows to compensate the different attenuation of the emitted signals caused by different receiver-transmitter distances. 
For the RxTx modules, the emission frequency, level, duration and the delay between emissions can be adjusted. 
A schematic of the HFLBL system is shown in Figure \ref{schematic}.

\subsection{Determination of Distances and Positions \label{high2}}

The relative position of each element of the acoustic positioning system can be evaluated by analysing the data after corrections and cleaning steps.
First a  correction has to be applied to all measurements performed by a single sensor in order to compensate for the propagation delays, up to 4$\, \mu$s, of the clock synchronization signals in the optical fibre network of the apparatus.
Then a correction has to be applied on the arrival times measured by the receiving modules to correct for the delay of the filtering and detection electronics and to compensate that low amplitude signals cross the predefined threshold later than high amplitude signals.
 The detection delay as a function of the measured amplitude of the acoustic signal (walk correction) was determined during a test campaign in a water pool. It amounts to 140-180$\,\mu$s and was found to be independent of the frequency of the signal in the frequency range used.  The walk correction uncertainty correlated to gain instabilities was estimated to be smaller than 10$\,\mu$s r.m.s.  Given a speed of sound of about 1500$\,$m/s, this leads to a distance uncertainty of about 1.5$\,$cm.

After correction of the walk effect, filtering and averaging over the arrival times is performed to clean the data sample from spurious measurements due to noise.
For each travel time measurement $T_{AB,i}$ between a specific emitter A and a receiver B  the sliding average is calculated including $n/2$ preceding and $n/2$  successive measurements (typically $n=6$):
\begin{equation}
t_{AB,i} = \langle T_{AB,i} \rangle =\frac{1}{n+1} \sum_{m=i-n/2}^{i+n/2} T_{AB,m} 
\end{equation}
The averaging technique can be applied since the position of a hydrophone changes  slowly due to the sea current. If $| T_{AB,i} - \langle T_{AB,i} \rangle |$ exceeds a given value the data point $i$ is judged as spurious and removed from the data sample. This procedure is iterated several times with ever tighter cuts, typically at 160$\,\mu$s, 80$\,\mu$s and 40$\,\mu$s, in order to optimally reject spurious detections without removing correct measurements.

\begin{figure}[ht]
 \begin{center}
 \centerline{\psfig{figure=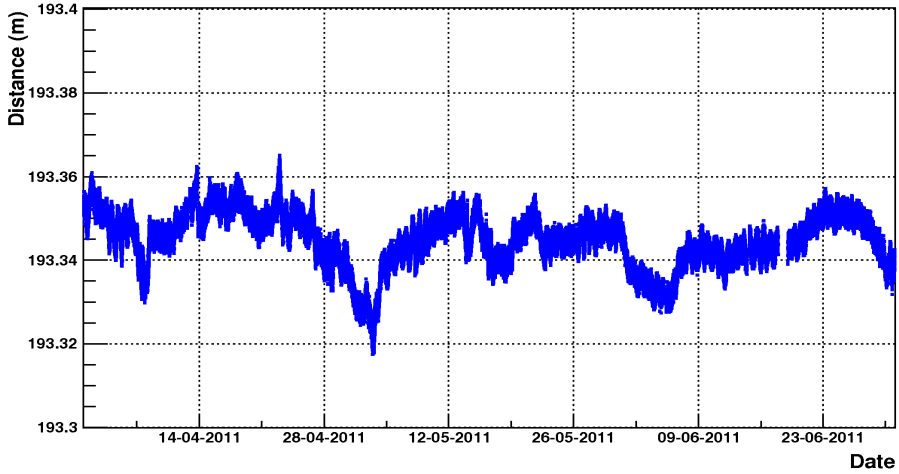,width=13.5cm,height=7.cm}}
 \end{center}
 \caption{Distance measured between two fixed RxTx hydrophones on two different BSS for a three month period in spring 2011.}  
\label{FixDist}
\end{figure}

The distance between the devices A and B can be calculated from the filtered travel times $t_{AB,i}$ knowing the speed of sound in sea water. As the speed of sound is not constant\footnote{The speed of sound in sea water depends on the temperature, salinity and the pressure (and thus on depth); the temperature and salinity values are almost constant over the entire ANTARES detector.} the travel time is defined as the integral over the path $s$ from A to B divided by the local sound velocity $c(s)$:
\begin{equation}
\label{path}
t_{AB,i} = \int_A^B \frac{ds}{c(s)}
\end{equation}
The ANTARES detector includes several sound velocimeters that determine, locally, the sound speed in water by measuring the travel time of an acoustic pulse over a distance of  20$\,$cm.
At the depth of the ANTARES neutrino telescope the speed of sound changes only with pressure i.e. with depth $z$ ($z$-axis chosen as vertically upward):
\begin{equation}
c(z)=c(z_0) - k_c (z-z_0)
\end{equation}
where $k_c=1.710$$\,\rm\frac{cm}{s} \frac{1}{m}$  derived from \cite{schall} for the Mediterranean Sea and 
$z_0$ is the depth of a sound velocimeter used to measure the speed of sound. The sound in deep-sea water does not propagate along a straight line, its trajectory is slightly bent. The bending radius of  about 80$\,$km \cite{Urick} is more than 100 times larger than  distances that have to be measured: for this reason the sound travel path is approximated as straight and the distances between acoustic modules are calculated as
\begin{equation}
d_{AB,i}=t_{AB,i} \left[ c(z_0) - k_c \left( \frac{z_A+z_B}{2} - z_0\right) \right]
\end{equation}
where $z_A, z_B$ are the depths of modules A and B. The error on the path length caused by this approximation is smaller than 2.5$\,$cm  for a distance of 500$\,$m.
The accuracy of the system is illustrated in Figure \ref{FixDist} which shows the distance measured between two fixed RxTx hydrophones during a period of three months. The plot exhibits a resolution of better than 1$\,$cm and a stability better than 3-4$\,$cm over a distance of 193$\,$m. The observed variations on the fixed distance are mainly due to the effect of changing sea current velocity along the acoustic wave travel path.

If during one measurement cycle at least three distances for a receiver module with respect to emitter modules of known positions are available, a three-dimensional triangulation of the module position is possible:
\begin{equation}
(\vec{r}_i - \vec{r}_j)^2 = d_{ij}^2
\end{equation}
with $\vec{r}_i$, $\vec{r}_j$ being the positions of acoustic modules $i$ and $j$, $d_{ij}$ the distance between module $i$ and $j$ calculated from the signal propagation time between $i$ and $j$ as described above.
Usually more than three distances to emitting modules are available which leads to an over-determined set of equations. The positions $\vec{r}_i$ are obtained by minimisation through an iterative convergence procedure based on Singular Value Decomposition of the linearised set of distance equations.

\begin{figure}[ht]
 \centering
\includegraphics[width=12.5cm,height=6.0cm]{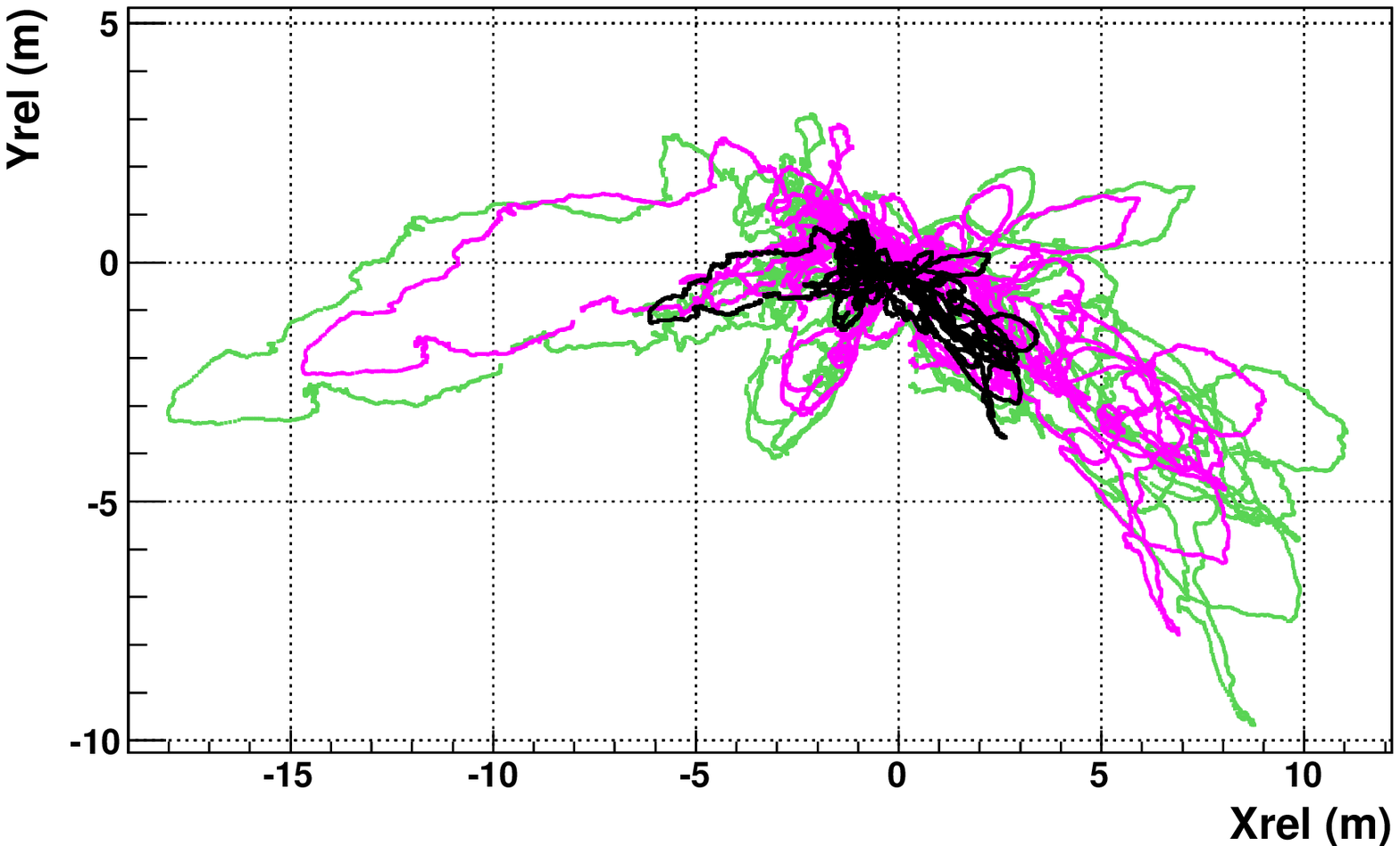}
\includegraphics[width=12.5cm]{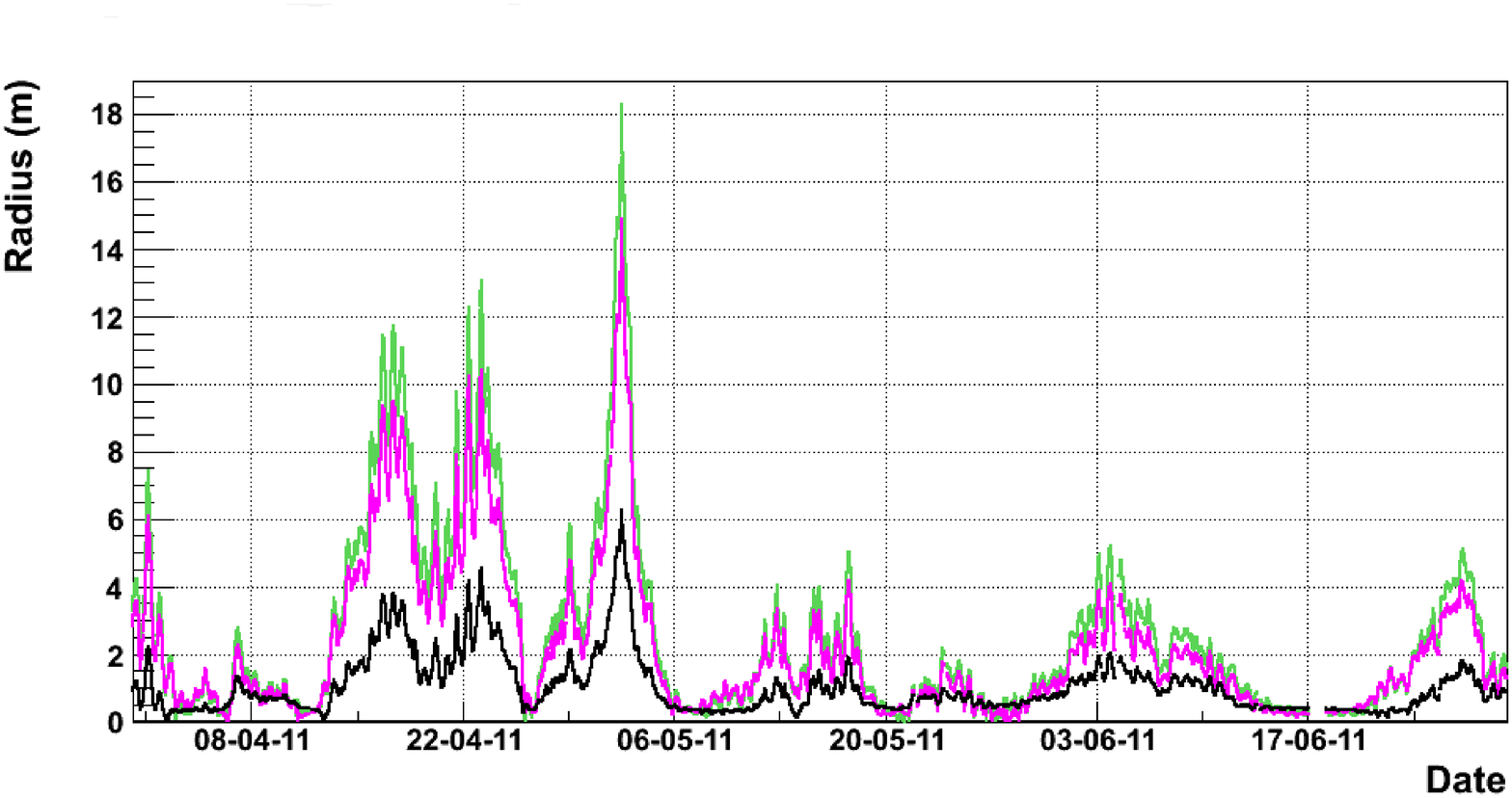}
 \caption{Top: Displacements of three hydrophones of Line 3 in the horizontal plane from April to June 2011 (black: storey 1, about 100$\,$m above sea bed; pink/dark grey: storey 14, about 290$\,$m above sea bed; green/light grey: storey 25, at the top of the line); Bottom: Time dependence of the radial displacement of these hydrophones for the same period.}
  \label{tria}
\end{figure}

An example for the hydrophone positions of one line obtained by triangulation is shown in Figure \nolinebreak\ref{tria}. The horizontal movement of three hydrophones at different altitudes along the line  and the absolute value of the displacement from the nominal position as a function of time are presented. The observed displacements increase with increasing height, reaching in this period up to 18$\,$m for the uppermost storey. Studies performed by Monte Carlo simulations taking into account the accuracy of the distances obtained from the acoustic travel times have demonstrated that the uncertainties of the hydrophone positions in the local detector reference frame vary  typically between 3 and 6$\,$cm from the lowest to the highest storey of the lines.

\section{The Compass-Tiltmeter System \label{CTS}}

Each  ANTARES storey is equipped with a commercial board (TCM2)\footnote{Manufactured by PNI Sensor Corporation, {\it http://www.pnicorp.com}} that measures the inclination of the storey with respect to the horizontal plane in two perpendicular axes (pitch and roll). The principle of this measurement is based on the movement of a fluid in the sensors due to the inclination of the storey. Since the inclinations measured with the tiltmeters give the tilts in the local frame of the TCM2 it is necessary to measure also the absolute orientation of the TCM2 (heading angle $\phi$, i.~e.~rotation around line axis). For this purpose three flux tube magnetic sensors measure the Earth magnetic field $\vec{B}$ in three perpendicular local directions ($x$, $y$, $z$) in the TCM2 frame.  
From the values $B_x$ and $B_y$ measured in the horizontal plane the heading angle $\phi$ between the $x$-axis of the TCM2 and the magnetic North  can be calculated:
\begin{equation}
\tan(\phi) = - \frac{B_y}{B_x}
\end{equation}
The angle with respect to the true North is obtained correcting the deviation between true and magnetic North which is 0.93$^{\circ}$ for January 2011 with an annual change by +0.12$^{\circ}$ \cite{mag}.
 The accuracy given by the manufacturer is 1$^{\circ}$ (resolution 0.1$^{\circ}$) for the heading angle and  0.2$^{\circ}$ (resolution 0.1$^{\circ}$) for the tilts.

\subsection{Calibration and Corrections}

\begin{figure}[ht]
 \centering
\includegraphics[width=6.0cm]{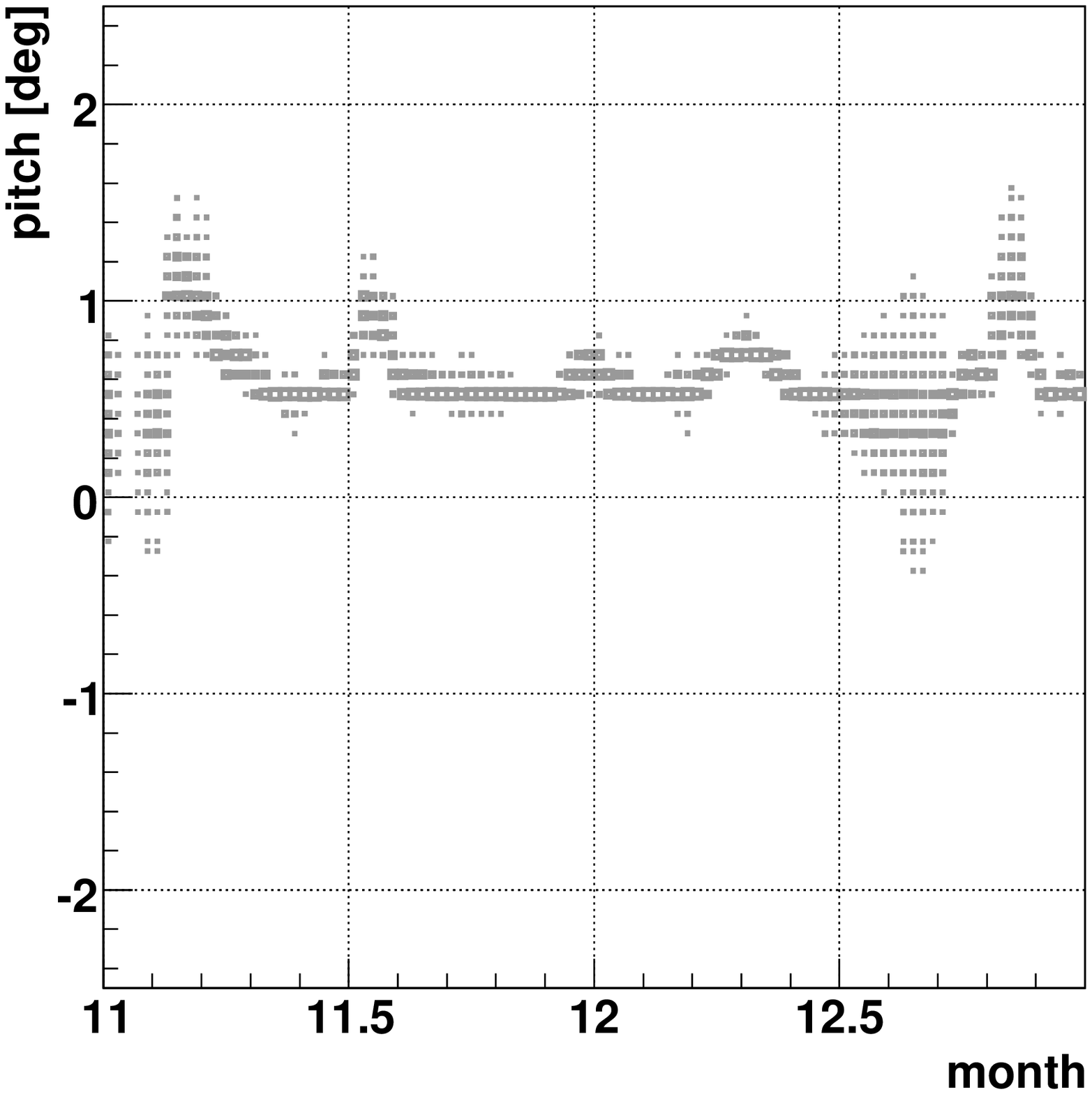}
\hspace*{5mm}
\includegraphics[width=6.0cm]{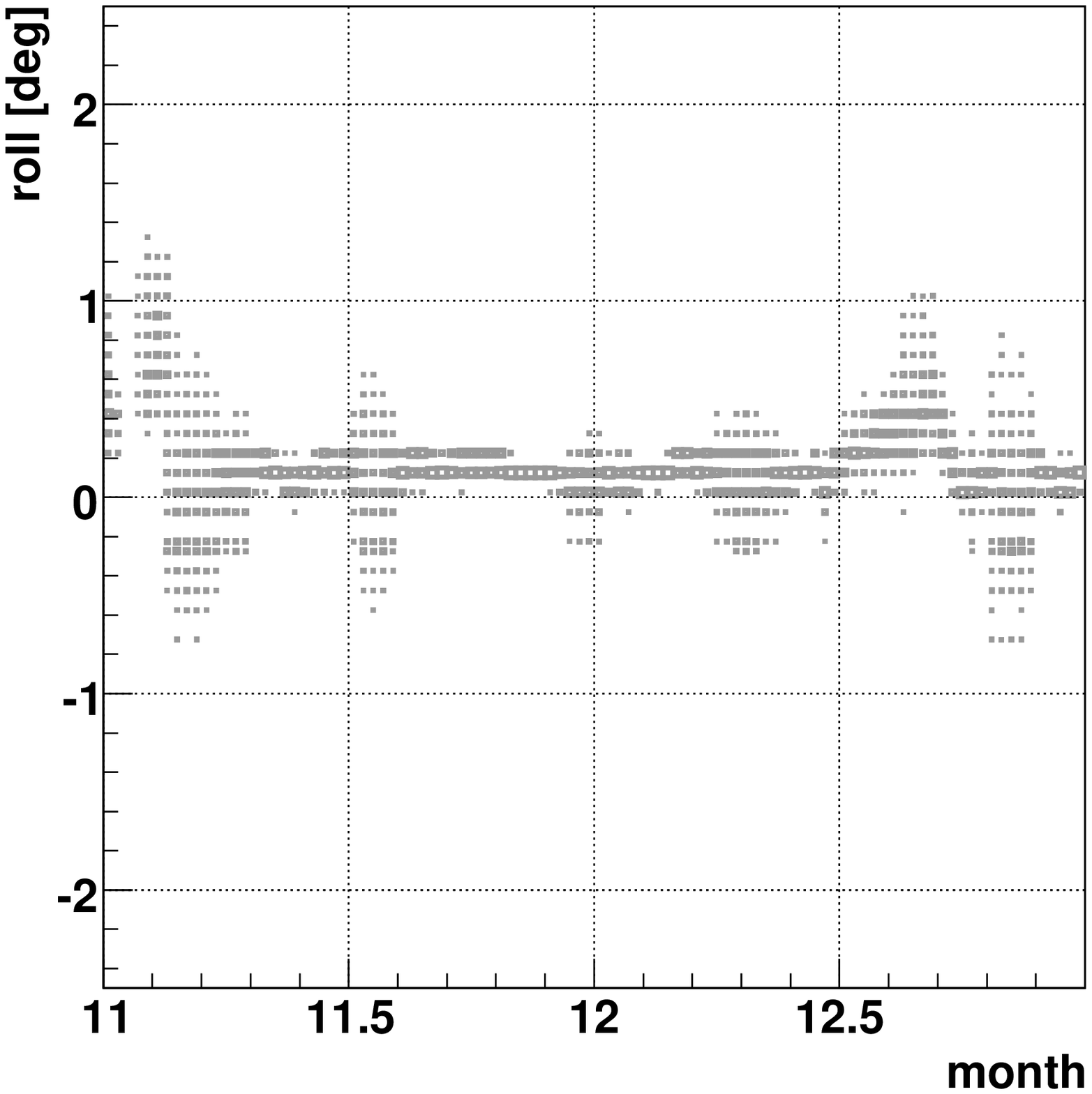}
 \caption{Measured pitch and roll values for Line 12 storey 9 in Nov./Dec.~2010 (for explanations see text).}
  \label{pitchroll}
\end{figure}

\noindent {\bf Tiltmeter Offsets:}
For typical values of the sea current (a few centimetres per second) the inclination of the storeys is of the order of 0.1$^{\circ}$. A precise measurement of the storey inclination requires an accurate knowledge on how the TCM2 card is mounted in the storey and of the intrinsic offsets of the TCM2 sensors. The orientation of the  TCM2 cards relative to the storeys have been measured during line construction.
  The offsets of the pitch and roll sensors were measured in the laboratory with the storeys positioned vertically. Typically the obtained offsets are in the range of 0.1-0.3$^{\circ}$,  so they have to be taken into account to avoid systematic errors in the reconstruction of the line inclination. A check of the tiltmeter offsets is performed {\it in situ} after line deployment: over a long period the values given by a tiltmeter are expected to fluctuate around the tiltmeter offset because the sea current direction changes leading to different orientations between the tiltmeters and the direction of the inclination of the storey. Figure \ref{pitchroll} shows the variation in time of the pitch and roll sensor of a certain storey over a two month period. The expected fluctuation around the offset is clearly visible and from the 'baseline' the offset can be determined (in this case 0.6$^{\circ}$ for pitch and 0.1$^{\circ}$ for roll).  

\begin{figure}[ht]
 \centering
\includegraphics[width=6.0cm]{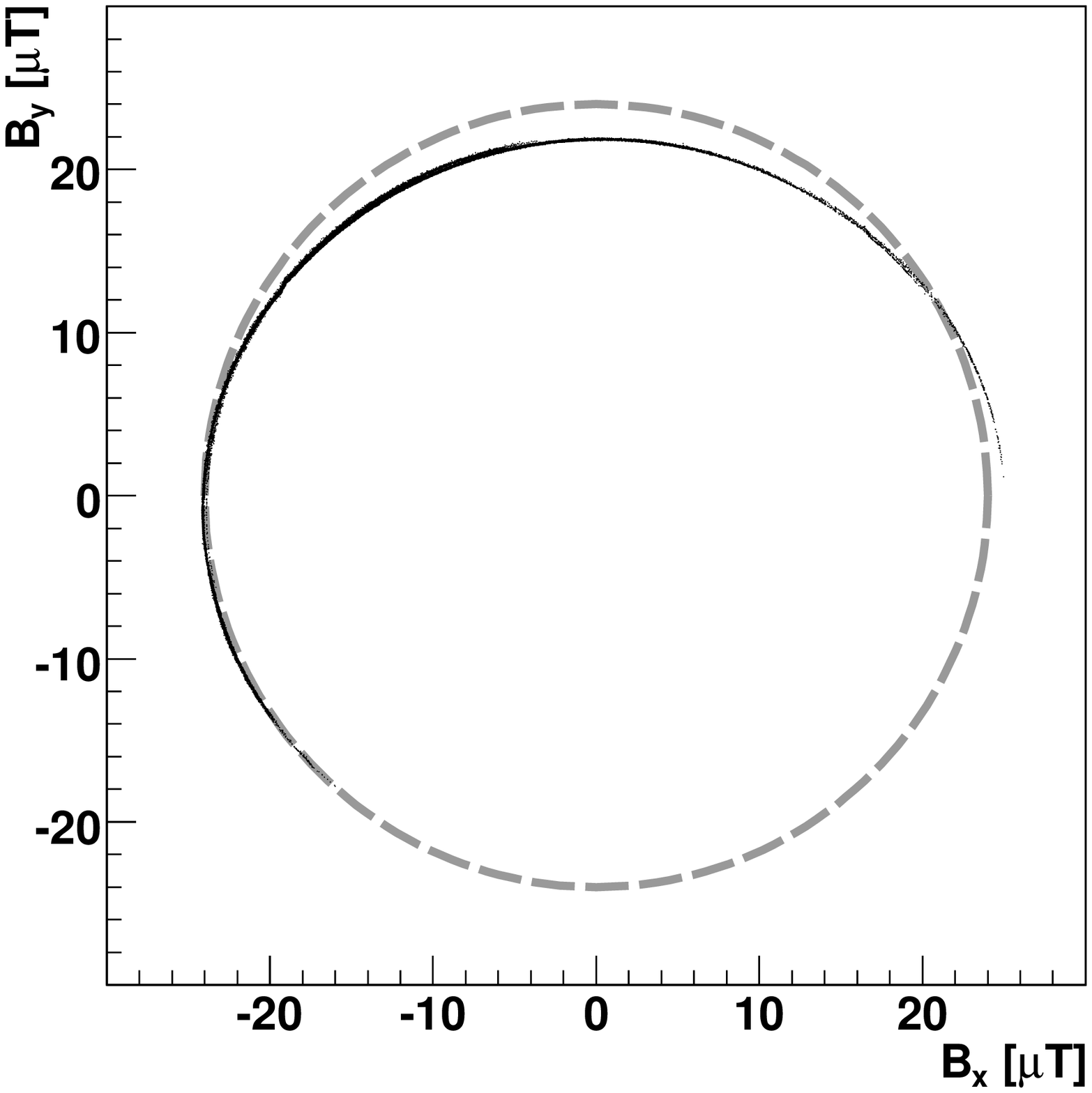}
\hspace*{5mm}
\includegraphics[width=6.0cm]{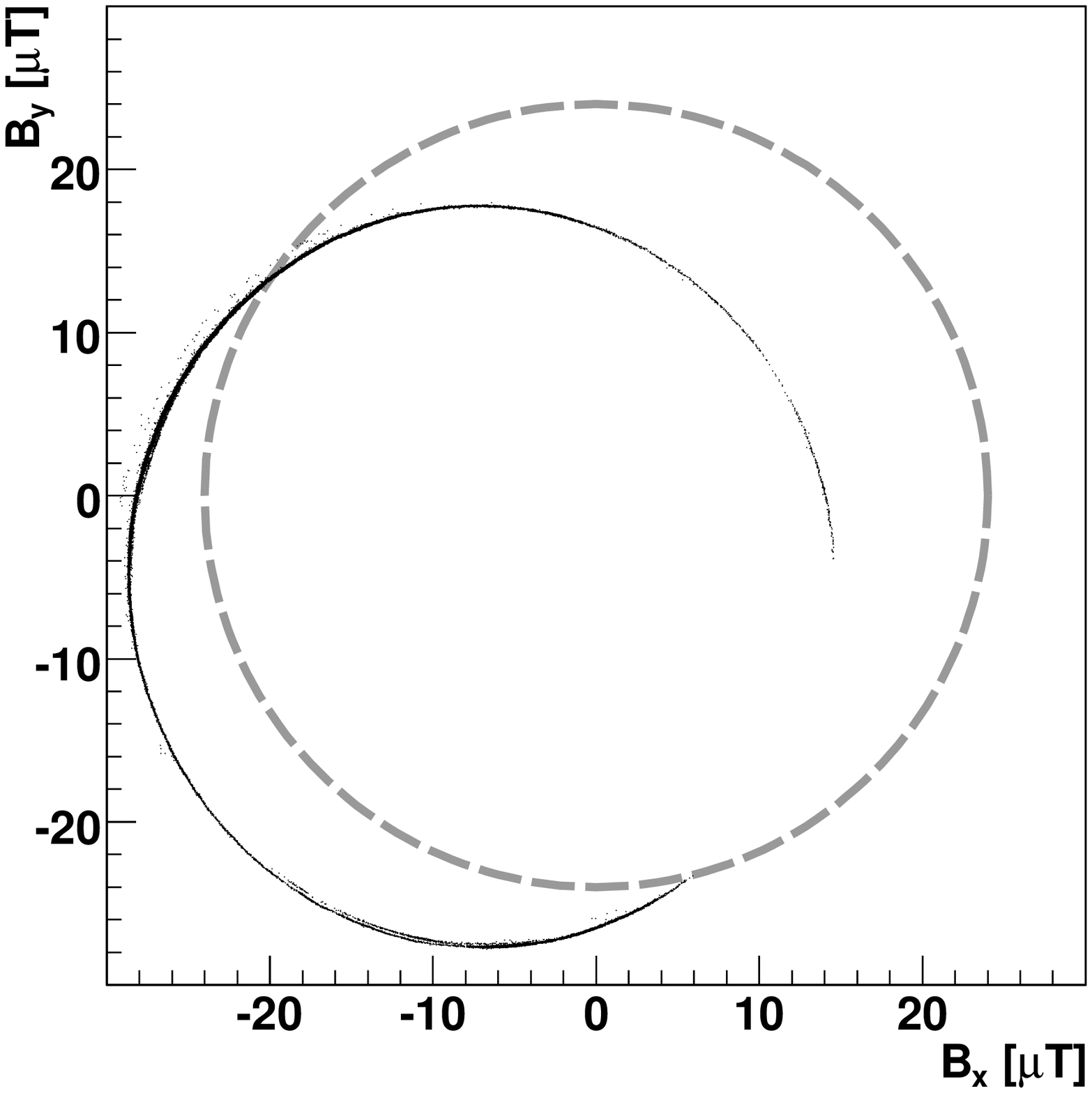}
 \caption{$B_y$ vs $B_x$ (black line) for two different storeys as measured in Nov./Dec.~2010 together with expected behaviour (dashed circle).}
  \label{bxyraw}
\end{figure}

\noindent {\bf Compass Calibration:}
The Earth magnetic field component in the horizontal plane at the ANTARES position (42$^{\circ}$48$^{\prime}$$\,$N, 6$^{\circ}$10$^{\prime}$$\,$E, elevation $-$2.4$\,$km), which is used to determine the heading of the storeys, amounts to 24.0$\,$$\rm\mu$T \cite{mag}.\footnote{The Earth magnetic field at the ANTARES site is mainly oriented North-South with only a small East-West component (values for Jan.~2011): $B_{\rm NS} = 24.0\,\rm\mu T$,  $B_{\rm EW} = 0.4\,\rm\mu T$ and $B_{z} = -39.4\,\rm\mu T$.} If a storey rotates 360 degrees  around the line axis in a dia\-gram of the measured $B_y$ vs $B_x$ a circle with radius 24.0$\,$$\rm\mu$T is expected. The change of the values of $B_y$ and $B_x$ due to the inclination of the storeys is negligible since the tilts are usually in the range of one degree or below. Figure \ref{bxyraw} shows  $B_y$ vs $B_x$ as observed in the deep-sea for storeys 13 and 25 of line 12, together with the expected behaviour. In some cases ellipses are observed instead of circles, in other cases the radius of the circles has not the expected value or the centre of the circles is shifted. 
The first two problems are  due to a miscalibration of the magnetic field sensors where $B_x$ and/or $B_y$ give too large or too small values; this can be corrected by a scaling factor $c_{x,y}$. The shift of the circles is due to parasitic magnetic fields inside the electronic container of the LCM, which follow the rotation of the storey.  Therefore they have a fixed direction in the local rotating frame of the LCM and are corrected by subtracting an offset from $B_x$ and $B_y$:
\begin{equation}
B_{x,y}^{\rm corrected}= c_{x,y} (B_{x,y}- B^{\rm off}_{x,y})
\end{equation}
The corrected values of  $B_x$ and $B_y$ are used to calculate the heading of the storey as described above. The correction for the heading of the storeys varies strongly from storey to storey and can be as large as 5 to 10 degrees. Thus for an accurate determination of the position and heading of the optical modules the application of this calibration is essential. The calibration constants are determined from {\it in situ} data. A proper calibration in the laboratory is not feasible because of the anthropogenic magnetic fields which are absent in the deep sea. 

\section{The Line Shape Model \label{lssection}}

A direct measurement of the storey position is only possible for the  five storeys containing a hydrophone. In order to evaluate the position of all storeys of a line  a mechanical model  was developed. The inclination  $\theta$ (zenith angle) of the line with respect to the vertical direction, at a certain height, is given by the ratio of the horizontal and vertical forces summed over all line elements $j$ above that point:
\begin{equation}
\tan (\theta_i) =\frac{\sum\limits_{j=i}^N F_j}{\sum\limits_{j=i}^N W_j}
\end{equation}
 The horizontal forces $F_j$ result from the flow resistance (or drag) of the line elements and are assumed to be parallel: there are contributions of the storeys, the cables in between (12.5$\,$m length), the cable from the BSS to the first storey (about 100$\,$m length)  and the buoy. The vertical forces $W_j$ are given by the difference of the buoyancy and the weight in air of each element. 
The flow resistance is calculated according to the following equation
\begin{equation}
\label{flow}
F_j =\frac{1}{2} c_{w,j} A_j \rho v^2
\end{equation}
 making use of the drag coefficients $c_{w,j}$ of the individual elements which were either measured or calculated; $\rho$, $A_j$ and $v$ are the density of the fluid (sea water), the cross section of the element and the sea current velocity, respectively\footnote{The vertical component of the current was found to be very small and is neglected.}. 
It is convenient to set up a continuous function by smearing the effective weight of the storeys and cables as well as the flow resistances over the whole line of height $h$. This will typically give:
\begin{equation}
W(z)=\left[ 25 \left(W_{\rm storey} + W_{\rm cable12m}\right) +W_{\rm cable100m} \right]\frac{h-z}{h} + W_{\rm buoy}
\end{equation}
and
\begin{equation}
F(z)=\left[ 25 \left(f_{\rm storey} + f_{\rm cable12m}\right) + f_{\rm cable100m} \right] \frac{h-z}{h}v^2 + f_{\rm buoy}v^2
\end{equation}
The numerical values for the constants $f_j =\frac{1}{2} c_{w,j} A_j \rho$ and $W_j$ are given in Table \ref{tab1}.
\begin{table}
 \caption{Drag and weight constants used for the modelling of the ANTARES lines based on measurements and calculations; negative values indicate downward going forces.}
 \label{tab1}
\begin{center}
 \begin{tabular}{| c | c | c | c | c |} \hline
  Element $j$:  & storey & cable12m & cable100m & buoy \\ \hline  
 $ f_j $ & 383.8$\,$$\rm Ns^2$/$\rm m^2$  & 222$\,$$\rm Ns^2$/$\rm m^2$ & 1850$\,$$\rm Ns^2$/$\rm m^2$  & 453$\,$$\rm Ns^2$/$\rm m^2$ \\ \hline  
 $ W_j $  & 265.6$\,$N & -52.9$\,$N &  -440$\,$N  & $\approx$7$\,$kN \\ \hline  
 \end{tabular}
\vspace{0.5cm}
\end{center}  
\end{table}

\begin{figure}[ht]
 \begin{center}
  \centerline{\psfig{figure=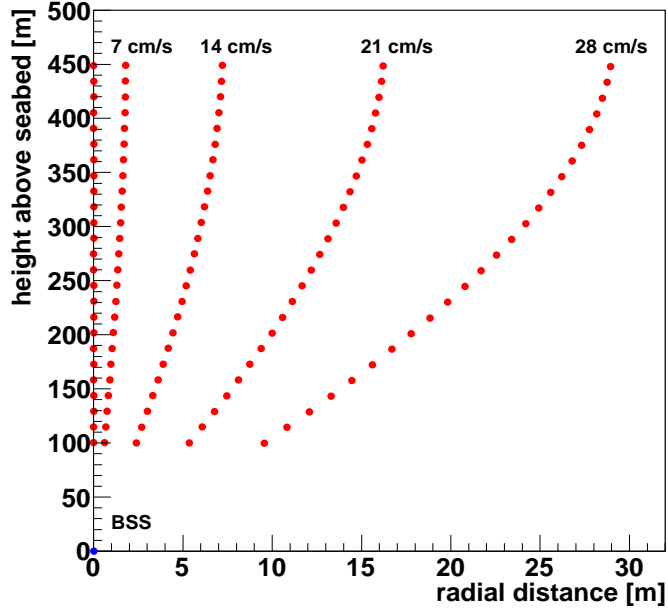,width=9.5cm,height=9.0cm}}
 \end{center}
 \caption{Calculated positions (height and radial displacement) of the storeys of a line (dots) with respect to the anchor of the line (BSS) for different sea current velocities according to the described line shape model; note the different scales on the axes.}
  \label{linev}
\end{figure}

Identifying $\tan\theta$ with the slope $dr$/$dz$ (with $r$ radial displacement at height $z$) of the line yields:
\begin{equation}
\label{drdz}
\frac{dr}{dz}=\frac{F(z)}{W(z)} = \frac{a-bz}{c-dz}v^2 = g(z)
\end{equation} 
with
\begin{equation}
a =  25 \left(f_{\rm storey} + f_{\rm cable12m}\right)  + f_{\rm cable100m} + f_{\rm buoy}, 
\end{equation}
\begin{equation}
b =  \frac{25}{h} \left(f_{\rm storey} + f_{\rm cable12m}\right)  +  \frac{1}{h}f_{\rm cable100m},  
\end{equation}
\begin{equation}
c =  25 (W_{\rm storey} + W_{\rm cable12m}) + W_{\rm cable100m} + W_{\rm buoy},
\end{equation}
\begin{equation}
d = \frac{25}{h} (W_{\rm storey} + W_{\rm cable12m})  +\frac{1}{h} W_{\rm cable100m}.
\end{equation}
By integration of equation (\ref{drdz}) over $z$, for a given value of the sea current $v$, the radial displacement $r(z)$ is obtained.
\begin{equation}
\label{lineshape}
r(z)= \int\limits_0^z g(z^{\prime}) dz^{\prime} = \left[ \frac{b}{d} z - \frac{ad-bc}{d^2} \ln\left(1 - \frac{d}{c} z \right) \right]v^2
\end{equation}
The combination of the linear and the logarithmic term is responsible for the characteristic shape of the line. The radial displacement increases quadratically with the sea current velocity.
Figure \ref{linev} illustrates the calculated positions of the 25 storeys of a line for different sea current velocities. For typical velocities (e.~g.~$v \le 7$$\,$cm/s, see Figure \ref{adcp}) the radial displacement is smaller than 2$\,$m even for the uppermost storey, whereas for a velocity of 20$\,$cm/s the displacement is as much as 15$\,$m. 

\section{Detector Alignment \label{LFIT}}

The line shape model described in the previous section is used to simultaneously fit the data of the acoustic positioning system and the data from the compass-tiltmeter system. The only free parameter is the current velocity:
the line shape equation (\ref{lineshape}) can directly be used to fit the position of the five hydrophones  from acoustic triangulation taking into account the known BSS position (anchor point of the line). Simultaneously the inclination angles of the 25 storeys of a line, which correspond to $\frac{dr}{dz}$, are fit according to equation (\ref{drdz}). In practice, a two-dimensional $\chi^2$-fit is performed on the $x$ and $y$ components of equations (\ref{drdz}) and (\ref{lineshape}) with  $v_x$ and $v_y$ (components of the sea current velocity in the horizontal plane) as fit parameters. This yields in addition to the radial displacement $r$ also the direction of the line inclination $\Psi$:
\begin{equation}
\tan\Psi = \frac{v_y}{v_x}
\end{equation}
For the calculation of the $\chi^2$ the respective Gaussian errors are taken into account, namely: 5$\,$cm for positions from triangulation, 1 degree for heading angles and 0.2 degrees for pitch and roll. 
From the radial displacement and the line inclination the three-dimensional positions of all storeys are calculated. Every two minutes, updates of the positions and orientations of the optical modules are thus available. The position of the centre of each storey and three angles describing the orientation of the storey  are stored in a database, together with the errors on the respective quantities obtained from error propagation. This procedure provides the complete geometry of the detector required for muon track reconstruction.

\begin{figure}[ht]
 \begin{center}
  \centerline{\psfig{figure=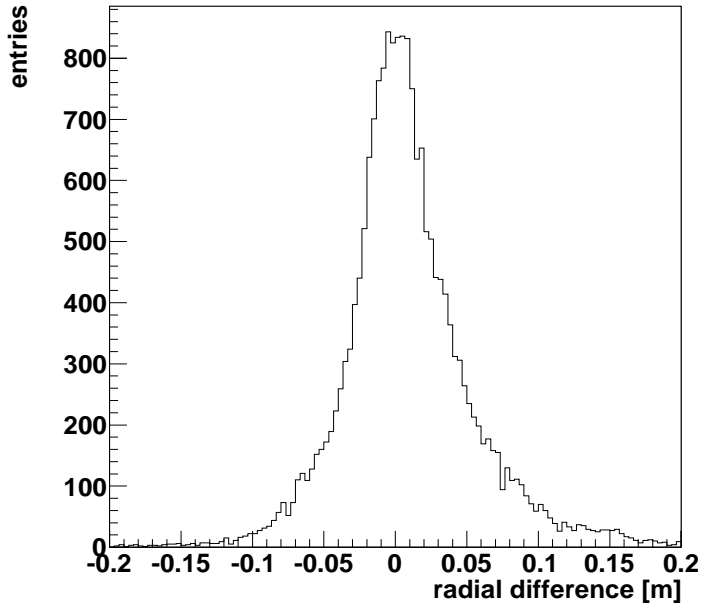,width=9.5cm,height=9.0cm}}
 \end{center}
 \caption{Difference between triangulation result and line fit for the position of storey 20 of Line 3 in March 2010.}
  \label{hydcomp}
\end{figure}

Figure \ref{hydcomp} shows the difference between the triangulated positions of a storey using its hydrophone and the positions resulting from the alignment fit for the periods in March 2010. A narrow distribution peaked almost at zero (mean: 0.8$\,$cm) is observed with an r.m.s. of 4.5$\,$cm. This excellent agreement indicates the
 absence of significant  systematic errors introduced by the line shape model and clearly demonstrates that the required precision of better than 10$\,$cm has been achieved.

\begin{figure}[ht]
 \begin{center}
  \psfig{figure=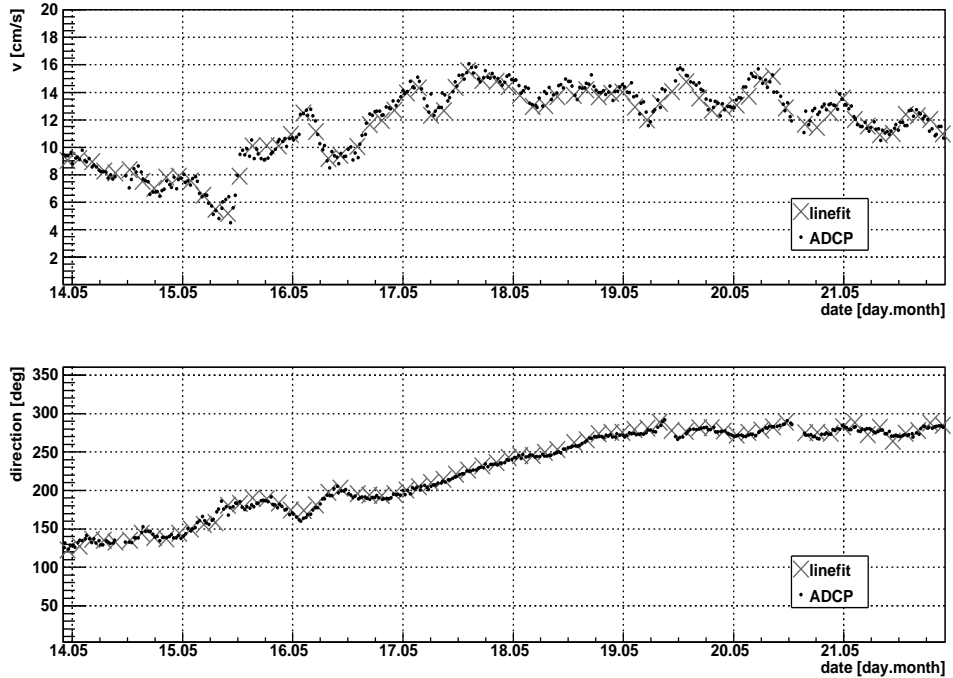,width=13.0cm,height=9.5cm}
 \end{center}
 \caption{Fitted current velocities from the positioning procedure (averaged over all twelve lines) compared to the ones measured with the Acoustic Doppler Current Profiler (ADCP) for a period of one week; top: speed of the sea current, bottom: direction of the sea current.}
  \label{adcplinefit}
\end{figure}

An additional validation of the line shape model and the fitting procedure is given by comparing the sea current velocity (absolute speed and direction values) obtained by the fit and the ones measured using the Acoustic Doppler Current Profiler. Over a broad range of velocities (5-15$\,$cm/s) there is good agreement between the measurements and the fit results (see Figure \ref{adcplinefit}). Given this agreement, the line fit procedure provides, as a byproduct, a continuous accurate long-term monitoring of the currents in the deep sea  every two minutes. Furthermore, the velocities obtained from the fit on the twelve individual lines agree well within their uncertainties (deviations typically less than 1$\,$cm/s). 

\section{Summary and Conclusion}

The positioning system of the ANTARES neutrino telescope, necessary to evaluate the position of the 885 optical modules with an accuracy better than 10$\,$cm, has been described with special emphasis on the applied corrections and cali\-brations.  The procedures used to filter and smooth the  arrival times for the acoustic triangulation, to correct for walk effects,  to account for the varying sound velocity with depth, to correct the offsets of the tiltmeters and to calibrate {\it in situ} the magnetic field sensors for the measurement of the heading of the optical modules, were described. 

The system has been in stable operation in its final configuration since summer 2008. It provides not only a full detector alignment every two minutes but gives -- in combination with a mechanical model for the line movement -- an accurate long-term monitoring of the sea current in the deep sea off the coast of Toulon.

The combined approach of an acoustic positioning system, for the triangulation of distances, and of a compass-tiltmeter system, for the measurement of inclinations and orientations, gives unambiguous information about the movement of the lines.  The resulting positions and orientations of the optical modules well match the requirements for track reconstruction with an underwater neutrino telescope like ANTARES. Further, the excellent performance of this system makes it a recommended technique for the  detector positioning for the planned future cubic kilometre scale neutrino telescope KM3NeT in the Mediterranean Sea \cite{KM3}. 

\section{Acknowledgements}

The authors acknowledge the financial support of the funding agencies: 
Centre National de la Recherche Scientifique (CNRS), Commissariat 
\`a l'\'ene\-gie atomique et aux \'energies alternatives  (CEA), Agence
National de la Recherche (ANR), Commission Europ\'eenne (FEDER fund 
and Marie Curie Program), R\'egion Alsace (contrat CPER), R\'egion 
Provence-Alpes-C\^ote d'Azur, D\'e\-par\-tement du Var and Ville de 
La Seyne-sur-Mer, France; Bundesministerium f\"ur Bildung und Forschung 
(BMBF), Germany; Istituto Nazionale di Fisica Nucleare (INFN), Italy; 
Stichting voor Fundamenteel Onderzoek der Materie (FOM), Nederlandse 
organisatie voor Wetenschappelijk Onderzoek (NWO), the Netherlands; 
Council of the President of the Russian Federation for young scientists 
and leading scientific schools supporting grants, Russia; National 
Authority for Scientific Research (ANCS), Romania; Ministerio de Ciencia 
e Innovaci\'on (MICINN), Prometeo of Generalitat Valenciana and MultiDark, 
Spain. We also acknowledge the technical support of Ifremer, AIM and 
Foselev Marine for the sea operation, the CC-IN2P3 for the computing facilities and the team of the former GENISEA company for their important contribution in the development of the acoustic positioning system.

\end{document}